\documentclass[twoside,11pt]{article}

\usepackage{authblk}
\usepackage{epsfig}
\usepackage{epstopdf}
\usepackage{graphicx}
\usepackage{color}
\usepackage{amsmath}

\newcommand{\udt}[3]{#1^{#2}_{\phantom{#2}#3}}

\newcommand{\dut}[3]{#1_{#2}^{\phantom{#2}#3}}

\begin{document}

\title{Some aspects of reconstruction using a scalar field in $f(T)$ Gravity}
\author[1]{Soumya Chakrabarti\footnote{email: soumya@cts.iitkgp.ernet.in}}
\affil[1]{Centre for Theoretical Studies, Indian Institute of Technology, Kharagpur, 721302, India}
\author[2]{Jackson Levi Said\footnote{email: jackson.said@um.edu.mt}}
\author[2]{Gabriel Farrugia\footnote{email: gabriel.farrugia.11@um.edu.mt}}
\affil[2]{Institute of Space Sciences and Astronomy, University of Malta, Msida, MSD 2080, Malta}
\affil[2]{Department of Physics, University of Malta, Msida, MSD 2080, Malta}

\maketitle

{\em PACS Nos. 04.50.Kd; 04.70.Bw
\par Keywords : $f(T)$ gravity, scalar field}
\vspace{0.5cm}

\begin{abstract}
General relativity characterizes gravity as a geometric property exhibited on spacetime by massive objects while teleparallel gravity achieves the same results, at the level of equations, by taking a torsional perspective of gravity. Similar to $f(R)$ theory, teleparallel gravity can also be generalized to $f(T)$, with the resulting field equations being inherently distinct from $f(R)$ gravity in that they are second order, while in the former case they turn out to be fourth order. In the present case, a minimally coupled scalar field is investigated in the $f(T)$ gravity context for several forms of the scalar field potential. A number of new $f(T)$ solutions are found for these potentials, with their respective state parameters also being examined.
\end{abstract}

\section{Introduction}
General Relativity is widely regarded as the fundamental theory of gravity which acts in accordance with certain well-defined foundation requirements and has passed a myriad of astrophysical tests (for a summary and systematic discussion of relevant issues we refer to the reviews of Faraoni Ref.\cite{Faraoni:2008mf}, Clifton, Ferreira, Padilla and Skordis \cite{Clifton:2011jh}). \medskip

However, there exists a tension between general relativity and observation that can only be eased by means of the imposition of exotic matter contributions. For instance, the recently observed accelerating expansion of the universe \cite{Riess:1998cb, Perlmutter:1998np} requires a dark energy contribution for general relativity to correctly describe late-time cosmology. Moreover, in the galactic regime, the velocity profile behavior compels the inclusion of the so-called dark matter component which originates from the efforts to explain the observed mismatch between the gravitational mass and the luminous mass of galaxies and clusters of galaxies. The gravitational mass of an object is determined by measuring the velocity and radius of the orbits of its satellites. The mass-to-light comparison indicates that the energy in luminous matter contributes less than $5$ percent of the average energy density of the universe. At lower scales without the dark matter proposition GR cannot produce the correct galactic rotation curves that are observed in galactic systems \cite{Zasov:2017gem}. Similarly, as the larger observational universe is compared with the predictions of GR, one requires the introduction of dark energy to reconcile observation with theory \cite{PhysRevD.60.081301, doi:10.1142/S0218271800000542, doi:10.1142/S021827180600942X}. \medskip

The simplest possible modification of GR is based on a cosmological constant. It carries its own complications, for example, the predicted energy scale is too large to be consistent with the vacuum energy predicted by quantum theory. A fairly popular choice is a scalar field with a slowly varying potential which serves as a competent candidate for inflation as well as for dark energy. Scalar fields have been used for a number of cosmological mechanisms such as inflation which can be constrained through CMB observations. Currently Planck data favors a single field slow-roll inflation \cite{Eshaghi:2015rta,Ade:2015lrj}. This is compatible with a number of modified theories of gravity. On the other hand, while scalar field models correspond to a modification of the energy-momentum tensor, there is another approach to be considered, which rests upon modifying the theory of gravity itself and as a consequence there is an effective energy-momentum tensor with a purely geometrical origin due to the modification of the Einstein-Hilbert action \cite{Sotiriou:2008rp, DeFelice:2010aj}. Rigorous attention has been given to theories where general relativity is modified by making the action a more general analytic function $f (R)$ of the Ricci scalar $R$. These are generically referred to
as $f (R)$ theories of gravity. The merits and demerits of these theories have been thoroughly studied in literature (for detailed discussions see Refs.\cite{Sotiriou:2008rp, Clifton:2011jh}). There are well defined criteria for viability of $f(R)$ theories of gravity which one must take care of with additional caution. This generates a problem regarding the adaptation of one particular form of $f(R)$ that can in principle work for the entire history of the universe. Moreover, the field equations of $f(R)$ gravity impose a second problem in the sense that they turn out to be fourth order in metric components and therefore become extremely difficult to inspect, for many scenarios of interest. \medskip

One proposition that has gained increasing interest in recent years is that of the teleparallel gravity \cite{Cai:2015emx, Krssak:2015oua, Hayashi:1979qx, Paliathanasis:2016vsw}. The theory finds its origin with Einstein himself, who first used the concept to work on the reunification of gravity with electromagnetism \cite{2005physics...3046U}. He attempted to do this by constructing an equivalent reformulation of GR. The motivation was to formulate an exact reproduction of GR at the level of equations. The difference between the two theories lies in the mechanism by which gravity is characterized \cite{Aldrovandi:2013wha}. Taking the Einstein-Hilbert action, gravity materializes by means of curvature on the manifold while in teleparallelism gravity is characterized as a torsional quantity. \medskip

Through a detailed covariant formulation of $f(T)$ gravity, the teleparallel lagrangian can also be generalized as was discussed in Refs.\cite{Krssak:2015oua}. Throughout this work, we consider this formulation of the theory. An important difference between $f(R)$ gravity and this new approach to gravitation is that while the $f(R)$ field equations are fourth order in terms of metric components, in the current generalization of the equivalent teleparallel formulation, the field equations continue to be second order. Therefore, at the very least the resulting equations turn out to be more tractable, and distinct from the $f(R)$ case. \medskip

The goal of the present work is to investigate the field equations of $f(T)$ gravity in the presence of a time-dependent self-interacting scalar field minimally coupled to gravity, which defines the matter contribution. No functional form of $f(T)$ is assumed at the outset. The main motivation of such an investigation is to reconstruct the exact form of the teleparralel lagrangian, i.e., to write $f(T)$ explicitly as a function of $T$ by solving  the field equations exactly. A number of useful solutions are found. For some cases the functional form of $f(T)$ are too complicated for any further investigations, however, a few very simple and relevant examples are reconstructed. Moreover, since the cosmological solutions found are simple to work with, we consider several forms of the self-interacting scalar field potential with the goal of reconstructing the scale factor, cosmological parameters with the main focus being defining the underlying Lagrangian that gives rise to that type of scalar field. \medskip

In general, scalar fields arise naturally in particle physics and string theory among other areas. For instance scalar fields arise in the low-energy limit of higher-dimensional theories \cite{BARROW1988515, WHITT1984176, Wands:1993uu}. They can also mimic the evolution of many kinds of matter, for instance it can be proved that the collapse of a spherically symmetric matter distribution can be modelled with power-law scalar field potentials \cite{Goncalves:1997qp}. Quadratic potentials for instance mimic pressure-less dust collapse whereas quartic potentials mimic radiation-like behavior \cite{Sami2007}. Very recently, dynamics of a self-interacting massive scalar field was discussed in Ref.\cite{Chakrabarti:2016xbk, Banerjee:2017njk}, under the assumption of conformal flatness, without any a priori choice of equation of state, exploiting the integrability of the scalar field evolution equation straightaway. The additional relevance of such investigations stems from the present importance of a scalar field to take on the role of the so-called dark energy, the agent responsible for the late-time acceleration of the universe whose distribution remains an open aspect of investigation. The key methodology of the study involved a detailed analytical investigation of the integrability of the Klein-Gordon equation governing the scalar field evolution, which can indeed be identified of a special case of an anharmonic oscillator equation. \medskip

Unless stated otherwise, geometric units are used where $G=1=c$. Also Latin indices are used to refer to local inertial coordinates while Greek ones are used to refer to global coordinates. The paper is divided as follows, in section II, $f(T)$ gravity is introduced with some focus on its cosmological effects. In section III, an important theorem of the Klein-Gordon equation is presented. This will help in determining solutions to some of the potential functions that will be considered. The different functional forms of the scalar field potential are then considered with the reconstruction work all contained in section IV. Finally the results are summarized and discussed in section V.

\section{$f(T)$ Gravity and Cosmology}
As with GR, its modifications and variants primarily depend on the metric tensor, $g_{\mu\nu}$, which is the fundamental dynamical object. This tensor is not dissimilar to a potential quantity, whereas spacetime curvature is exhibited through the Levi-Civita connection (torsion-free), $\Gamma^{\lambda}_{\mu\nu}$. On the other hand, teleparallelism replaces this connection with the Weitzenb\"{o}ck connection, $\hat{\Gamma}^{\lambda}_{\mu\nu}$, which is curvature-free. This new connection is determined by two dynamical variables, namely the tetrads (or vierbein) and the spin connection. \medskip

The tetrads, $\udt{e}{a}{\mu}$, relate inertial and global coordinates, that is, they relate the Minkowski metric, $\eta_{\mu\nu}$, with the metric tensor. They can be used to relate physical observers with the spacetime under consideration. Therefore, the metric tensor can be represented by \cite{Paliathanasis:2016vsw}
\begin{equation}
g_{\mu\nu}=\eta_{ab}\udt{e}{a}{\mu}\udt{e}{b}{\nu}.
\end{equation}
where $\eta_{ab}=\text{diag}(1,-1,-1,-1)$. The tetrads must satisfy the following inverse relations
\begin{equation}
\udt{e}{a}{\mu}\dut{e}{a}{\nu}=\delta^{\nu}_{\mu} \quad \udt{e}{a}{\mu}\dut{e}{b}{\mu}=\delta^{a}_{b},
\end{equation}
but are otherwise not unique for a given metric tensor. \medskip

The degree of freedom present in the choice of the form of the tetrad is countered by the spin connection, $\udt{\omega}{b}{a\mu}$, which is not a tensor and heavily dependent on the particular system under consideration \cite{Krssak:2015oua}. In fact, it is the spin connection that preserves the property of covariance in $f(T)$ gravity. Those tetrads that produce a vanishing spin connection are called {\it pure} tetrads while those that are nonzero are called {\it impure} tetrads \cite{Tamanini:2012hg}. \medskip

\noindent In the current work, we consider the flat FRW metric
\begin{align}
ds^2&=dt^2-a^2(t)\Sigma_{i=1}^{3}dx_i^2\nonumber\\
&=dt^2-a(t)^2(dx^2+dy^2+dz^2),
\label{metric}
\end{align}
where $x_i$ represents Cartesian coordinates. The natural choice of tetrad for this metric is
\begin{equation}
\udt{e}{a}{\mu}=diag(1,a(t),a(t),a(t)).
\label{frw_tetrad}
\end{equation}
This form of the FRW metric is very advantageous because its spin connection vanishes \cite{Krssak:2015oua}, and so no extra contribution is needed in the $f(T)$ field equations. \medskip

\noindent Consequently, the Weitzenb\"{o}ck connection takes on the form \cite{Hayashi:1979qx}
\begin{equation}
\hat{\Gamma}^{\lambda}_{\mu\nu}=\dut{e}{a}{\lambda}\partial_{\mu}\udt{e}{a}{\nu},
\end{equation}
which naturally leads to the torsion tensor definition \cite{Cai:2015emx}
\begin{equation}
\udt{T}{\lambda}{\mu\nu} \equiv \hat{\Gamma}^{\lambda}_{\mu\nu}-\hat{\Gamma}^{\lambda}_{\nu\mu}.
\end{equation} \medskip

The difference between GR and teleparallel gravity is characterized through the contorsion tensor which is defined as
\begin{equation}
\udt{K}{\mu\nu}{a} \equiv \frac{1}{2}\left(\dut{T}{a}{\mu\nu}+\udt{T}{\nu\mu}{a}-\udt{T}{\mu\nu}{a}\right).
\end{equation}
On a superficial level, the contorsion tensor illustrates clearly the differences at the level of the connection. \medskip

\noindent Finally, the superpotential is defined as
\begin{equation}
\dut{S}{a}{\mu\nu} \equiv \frac{1}{2}\left(\udt{K}{\mu\nu}{a}-\dut{e}{a}{\nu}\udt{T}{\alpha\mu}{\alpha}+\dut{e}{a}{\mu}\udt{T}{\alpha\nu}{\alpha}\right).
\end{equation}
This tensor does not expose any of the fundamental mechanisms of $f(T)$ gravity but serves to simplify a number of equations in the theory. \medskip

Analogous to the Ricci curvature scalar, the torsion scalar, $T=\udt{T}{a}{\mu\nu}\dut{S}{a}{\mu\nu}$, plays the role of Lagrangian for the teleparallel equivalent of general relativity (TEGR) case. In this setting, GR is completely reproduced at the level of equations. For the tetrad in Eq.(\ref{frw_tetrad}), the torsion scalar turns out to be \cite{Krssak:2015oua}
\begin{equation}
T=-6\frac{\dot{a}^2(t)}{a^2(t)}\equiv -6H^2.
\label{torsion_scalar}
\end{equation}

\noindent Following the same reasoning as $f(R)$ gravity, the action of TEGR can be generalized to
\begin{equation}
S=\frac{1}{4\kappa}\displaystyle\int d^4 x\,e \left(T + f(T)\right)+\displaystyle\int d^4 x\,e\mathcal{L}_m,
\label{torsion_action}
\end{equation}
where $\kappa=4\pi G$ and $e=\det{\udt{e}{a}{\mu}}$. TEGR is recovered on the choice of $f(T)=0$. On taking a variation with respect to the tetrad the following field equations emerge \cite{Cai:2015emx}
\begin{align}
& e^{-1} f_{T} \partial_{\nu}\left(e \dut{S}{a}{\mu\nu}\right)+f_{TT} \dut{S}{a}{\mu\nu} \partial_{\nu} T\nonumber\\
&-f_{T} \udt{T}{b}{\nu a}\dut{S}{b}{\nu\mu}+\frac{1}{4}f(T)\dut{e}{a}{\mu}=\kappa \dut{\Theta}{a}{\mu},
\end{align}
where $\dut{\Theta}{a}{\mu}\equiv \frac{1}{e}\frac{\delta \mathcal{L}_m}{\delta \udt{e}{a}{\mu}}$, $f_{T}$ and $f_{TT}$ denote the first and second derivatives of $f(T)$ with respect to $T$, and $\mathcal{L}_m$ is the matter lagrangian. \medskip

The matter contribution is assumed to be a spatially homogeneous scalar field, $\phi$, that is minimally coupled to the gravitational action with a scalar potential, $V(\phi)$. Therefore, the energy--momentum tensor turns out to be
\begin{equation}
T^\phi_{\mu\nu}=\partial_\mu\phi\partial_\nu\phi-g_{\mu\nu}\Bigg[\frac{1}{2}g^{\alpha\beta}\partial_\alpha\phi\partial_\beta\phi-V(\phi)\Bigg].
\label{scalarcontribution}
\end{equation}

\noindent By varying the action and taking this energy--momentum tensor results in the Friedmann equations
\begin{equation}
f - T - 2 T \frac{df}{dT} = \dot{\phi}^2 + 2 V(\phi),
\label{fe1}
\end{equation}
and
\begin{equation}
\dot{H} = - \frac{\dot{\phi}^2}{(1+\frac{df}{dT}+2T\frac{d^{2}f}{dT^{2}})},
\label{fe2}
\end{equation}
where dots denote derivatives with respect to cosmic time. \medskip

With these equations in hand, the action can also be varied with respect to the scalar field itself, giving the evolution equation
\begin{equation}
\ddot{\phi} + 3 H \dot{\phi} + \frac{dV}{d\phi} = 0.
\label{wave}
\end{equation}

The evolution equation can be immediately utilized to gain some insight into the scalar field. This will be investigated in the next section by using an integrability condition for an anharmonic oscillator equation.

\section{A note on the integrability of anharmonic oscillator equation}
The simple linear harmonic oscillator has a straightforward sinusoidal solution however the anharmonic oscillator incorporates more terms and can represent more physical features of a system. This takes the form of a nonlinear second order differential equation \cite{euler1997transformation}
\begin{equation}
\label{gen}
\ddot{\phi}+f_1(t)\dot{\phi}+ f_2(t)\phi+f_3(t)\phi^n=0,
\end{equation}
where $f_i$ are functions of $t$ and $n \in {\cal Q}$ is a constant. Overhead dot represents differentiation with respect to cosmic time, $t$. Using Euler's theorem on the integrability of the general anharmonic oscillator equation \cite{euler1997transformation} and the more applicable result by Ref.\cite{harko2013}, this equation can be integrated under certain conditions. Formally, the theorem can be stated in the following manner. \medskip

\textbf{Theorem} An equation of the form Eq.(\ref{gen}) can be transformed into an integrable form for $n\notin \left\{-3,-1,0,1\right\} $, provided the coefficients of Eq. (\ref{gen}) satisfy the differential condition
\begin{equation}
\label{int-gen}
\frac{1}{n+3}\frac{1}{f_{3}(t)}\frac{d^{2}f_{3}}{dt^{2}%
}-\frac{n+4}{\left( n+3\right) ^{2}}\left[ \frac{1}{f_{3}(t)}\frac{df_{3}}{dt%
}\right] ^{2}+ \frac{n-1}{\left( n+3\right) ^{2}}\left[ \frac{1}{f_{3}(t)}%
\frac{df_{3}}{dt}\right] f_{1}\left( t\right) + \frac{2}{n+3}\frac{df_{1}}{dt}%
+\frac{2\left( n+1\right) }{\left( n+3\right) ^{2}}f_{1}^{2}\left( t\right)=f_{2}(t).
\end{equation}

\noindent Introducing a pair of new variables $\Phi$ and $T$ given by 
\begin{eqnarray}
\label{Phi}
\Phi\left( T\right) &=&C\phi\left( t\right) f_{3}^{\frac{1}{n+3}}\left( t\right)
e^{\frac{2}{n+3}\int^{t}f_{1}\left( x \right) dx },\\
\label{T}
T\left( \phi,t\right) &=&C^{\frac{1-n}{2}}\int^{t}f_{3}^{\frac{2}{n+3}}\left(
\xi \right) e^{\left( \frac{1-n}{n+3}\right) \int^{\xi }f_{1}\left( x
\right) dx }d\xi ,\nonumber\\
\end{eqnarray}%
where $C$ is a constant, Eq.(\ref{gen}) can then be written in an integrable form as 
\begin{equation}
\label{Phi1}
\frac{d^{2}\Phi}{dT^{2}}+\Phi^{n}\left( T\right)=0.
\end{equation}

Using the transformation equations (\ref{Phi}) and (\ref{T}), one can write the general solution for the scalar field $\phi$ as,
\begin{equation}
\label{phigen}
\phi\left( t\right) =\phi_{0}\left[ C^{\frac{1-n}{2}}\int^{t}f_{3}^{\frac{2}{n+3}}\left( \xi \right) e^{\left( \frac{1-n}{n+3}\right) \int^{\xi }f_{1}\left(x \right) dx }d\xi -T_{0}\right] ^{\frac{2}{1-n}}f_{3}^{-\frac{1}{n+3}}\left( t\right) e^{-\frac{2}{n+3}\int^{t}f_{1}\left( x \right) dx},
\end{equation}
where $\phi_{0}$ and $T_0$ are constants of integration and $C$ comes from the definition of the point transformations (\ref{Phi}) and (\ref{T}). Both $\phi_{0}$ and $C$ must be non-zero. \medskip

In what follows, we shall use this integrability condition in order to extract information from the scalar field evolution equation as in Eq.(\ref{wave}) for some suitable forms of the potential $V = V(\phi)$. Apart from mathematical elegance and simplicity, the choice of the potential can indeed represent various physical characteristics which we will discuss for each form of $V(\phi)$.

\section{Cosmological Solutions for different self-interaction potentials}
In the following work we consider a number of different scalar field  potentials in order to reconstruct lagrangian functional forms for $f(T)$ with state parameters also being provided.

\subsection{Simple power law potential}
We assume the self-interaction to be a power law function of the scalar field such that $\frac{dV(\phi)}{\phi} = \phi^{n}$. For positive powers of $\phi$, the effective mass of the field can be determined by evaluating $\frac{d^{2}V(\phi)}{d\phi^{2}}$ at $\phi = 0$. Inverse powers of the field potential are also very useful in the cosmological setting, as quintessence fields among other interesting properties. One such example was studied by Peebles and Ratra \cite{1988ApJ...325L..17P} where the potential $V(\phi) = \frac{M^{(4+\alpha)}}{\phi^{\alpha}}$ was used with $M$ being the Planck mass. This model had some interesting effects on inflation within the context of GR with a minimally coupled scalar field. Similarly, other potential of this nature have also been investigated, for example by Steinhardt, Wang and Zlatev \cite{PhysRevD.59.123504} where the mass parameter is allowed to veer away from the Planck mass. In this case the authors investigate tracker solutions which allow for a larger range of initial conditions that eventually lead to similar late time cosmological behavior and include an inflationary epoch. \medskip

There are a large number of cosmological models that are based on scalar fields, each with their own way of mimicking the dark energy component of the universe. Another aspect of many of these models is their consequences for inflation. Given the lack of data on the inflationary epoch it is unclear how to compare the numerous models. However, some of these various potentials and solutions are discussed at length in Refs.\cite{NAKAYAMA2013111,2012IJMPD..2130002Y,Sami2007}. \medskip

For the potential in question here, we compare the resulting evolution equation using Eq.(\ref{wave}) with the general anharmonic oscillator equation given in the theorem in Eq.(\ref{gen}), and find that the coefficients take on the following form $f_{1} = 3\frac{\dot{a}}{a}$, $f_{2} = 0$, $f_{3} = 1$. This naturally leads to the criterion of integrability in Eq.(\ref{int-gen}) giving the second order non-linear differential equation
\begin{equation}
\dot{a}^2 a^{\frac{4n}{(n+3)}} = \lambda^2,
\end{equation}
which naturally results in the scale factor
\begin{equation}\label{exactscale1}
a(t) = [\lambda (t-t_{0})]^{\frac{(n+3)}{3(n+1)}}.
\end{equation}
In this solution it is assumed that $\dot{a} > 0$.

\begin{figure}[ht]
\begin{center}
\includegraphics[width=0.35\textwidth]{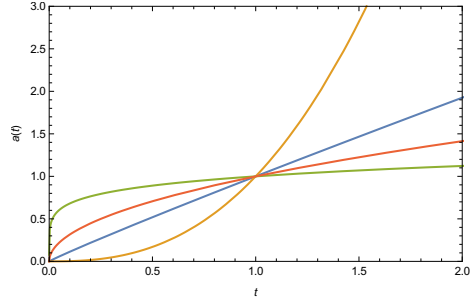}
\caption{Evolution of the scale factor with cosmic time for power-law potential. The various colors take on the following values of the index $n$ -- Red: $n = 3$, Blue: $n = 0.09$, Yellow: $n = -0.7$, and Green: $n = -5$.}
\label{power_law_graph}
\end{center}
\end{figure}

In Fig.(\ref{power_law_graph}), we plot the evolution of the scale factor with cosmic time for various values of the index $n$. We find that positive values of $n$ show behavior similar to a flat cosmology, while marginally negative values of $n$ resonate better with open cosmologies. Scale factor with negative powers are found for $-3<n<-1$ which are clearly unrealistic. The $-1 < n < 0$ region is also very interesting in that it clearly shows the possibility of late time expansion. \medskip

\begin{figure}[ht]
\begin{center}
\includegraphics[width=0.35\textwidth]{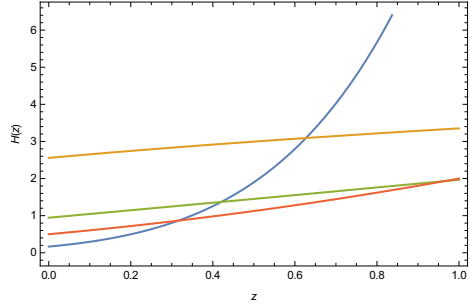}
\caption{Evolution of Hubble parameter with respect to redshift for the power-law potential. The corresponding values of the index $n$ are the same as in Fig.(\ref{power_law_graph})}
\label{Hubble_powerlaw}
\end{center}
\end{figure}

\begin{figure}[ht]
\begin{center}
\includegraphics[width=0.35\textwidth]{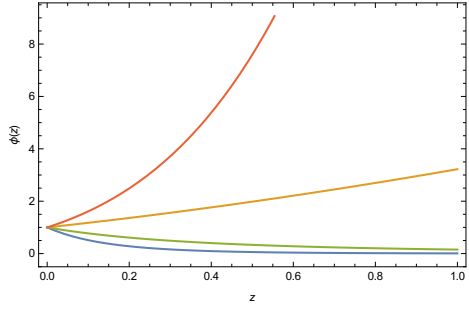}
\caption{Evolution of scalar field with respect to redshift for the power-law potential. The corresponding values of the index $n$ are the same as in Fig.(\ref{power_law_graph})}
\label{Scalar_powerlaw}
\end{center}
\end{figure}

\begin{figure}[ht]
\begin{center}
\includegraphics[width=0.35\textwidth]{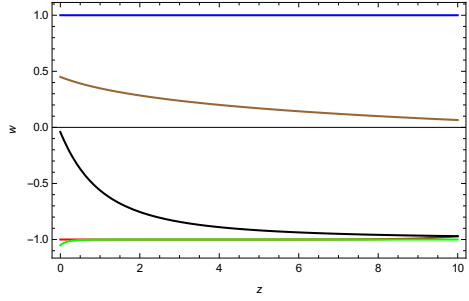}
\caption{Evolution of equation of state with respect to redshift for power-law potential; for different range of redshift. The various colors take on the following values of the index $n$ -- Red: $n = -1.5$, Blue: $n = -0.7$, Brown: $n = 0.09$, Black: $n = 0.5$, and Green: $n = 3$.}
\label{power_law_scalar}
\end{center}
\end{figure}

Given that an exact scale factor form is found through the integrability criterion, an explicit form of the scalar field can also be obtained using Eq.(\ref{phigen}). this results in
\begin{equation}\label{exactfield}
\phi(t) = \xi_{0} (t-t_{0})^{\frac{2(1+3n)}{(1-n^{2})}},
\end{equation}
where $\xi_{0}$ is a constant. \medskip

Using both the scale factor and scalar field we find the following ingredients of the Friedmann equation
\begin{equation}
\dot{\phi}^2 = \lambda_{0}^2 H^{\lambda_1},
\end{equation}
and
\begin{equation}
V(\phi) = \lambda_{2} H^{\lambda_3},
\end{equation}
where $n_{0}^2 = \frac{(n+3)}{3(n+1)}$, $\lambda_{0}^2 = \xi_{0}^2 \frac{4(1+3n)^2}{(1-n^2)^2} n_{0}^{\frac{4(1+n^{2}+6n)}{(1-n^2)}}$, $\lambda_1 = 2(n^{2}-1)$, $\lambda_2 = \frac{\xi_{0}^{(n+1)} n_{0}^{\frac{4(1+3n)}{(1-n)}}}{(n+1)}$, $\lambda_3 = \frac{4(1+3n)}{(n-1)}$. \medskip

Now taking the Friedmann equation in Eq.(\ref{fe1}), the functional form of $f(T)$ can be reconstructed. We do this in terms of the Hubble parameter, $H$, which results in the equation
\begin{equation}
f(H) + 6H^2 - H\frac{df}{dH} = \lambda_{0}^2 H^{\lambda_1} + \lambda_{2} H^{\lambda_3},
\end{equation} 
which can be rewritten as
\begin{equation}
\frac{d}{dH}\Big(\frac{f(H)}{H}\Big) = 6 - \lambda_{0}^2 H^{\lambda_{1} - 2} - \lambda_{2} H^{\lambda_{3} - 2}.
\end{equation}
Resulting in the lagrangian
\begin{equation}
f(H) = \lambda_{4} + 6 H^2 - \frac{\lambda_{0}^2}{(\lambda_{1}-1)} H^{\lambda_1} - \frac{\lambda_{2}}{\lambda_{3}-1} H^{\lambda_3}.
\end{equation}

\noindent Returning to the torsion scalar through Eq.(\ref{torsion_scalar}) gives the $f(T)$ lagragian as
\begin{equation}
f(T) = \lambda_{4} - T - \frac{\lambda_{0}^2}{(\lambda_{1}-1)} (-\frac{T}{6})^{(n^{2}-1)} - \frac{\lambda_{2}}{\lambda_{3}-1} (-\frac{T}{6})^{\frac{2(1+3n)}{(n-1)}}.
\end{equation}

\noindent With the scale factor in hand, we also investigate the deceleration parameter which takes on the form
\begin{equation}
q = -\frac{3(1+n)}{(3+n)}\Bigg(-1 + \frac{(3+n)}{3(1+n)}\Bigg).
\end{equation}

The Hubble parameter is shown in Fig.(\ref{Hubble_powerlaw}) as a function of the redshift. The result is a Hubble parameter continuously increasing with $z$ for all values of $n$. The inverse power potentials increase the least with redshift (the current case being $V(\phi) \sim \phi^{-4}$). The evolution of the scalar field is then shown in Fig.(\ref{Scalar_powerlaw}). Similarly, this maintains a positive profile along the various redshift values. On the other hand, the trend is decreasing for positive values of $n$. In these cases the scalar field asymptotically tends to zero. \medskip

\noindent The scalar field naturally leads to an EoS
\begin{equation}\label{EoS_eqn}
w = \frac{p_{\phi}}{\rho_{\phi}} = \frac{\frac{\dot{\phi}^2}{2} - V(\phi)}{\frac{\dot{\phi}^2}{2} + V(\phi)},
\end{equation}
which can easily be represented in terms of redshift for the various values of $n$. The resulting behavior is very sensitive to the choice of $n$, that is, whether it is positive or negative. The EoS naturally tends to $-1$ for some negative values of the potential index. The best behaved EoS factors are for $n=-1.5$ and $n=3$.

\subsection{Combination of Power-Law terms as a potential}
Quadratic potentials hold particular interest in scalar field theories due to its effect on inflation \cite{Linde:1983gd,Liddle:2000cg,Bassett:2005xm}. In this section we consider the case of a combined potential with the quadratic case. That is, we take as the effective potential to be
\begin{equation}
V(\phi) = \frac{1}{2}{\phi}^{2} + \frac{1}{n+1}\phi^{n+1},
\end{equation}
such that $\frac{dV}{d\phi} = \phi + \phi^n$. This gives a scalar field evolution that takes the form
\begin{equation}
\ddot{\phi} + 3\frac{\dot{a}}{a}\dot{\phi} + \phi + \phi^n = 0.
\end{equation}

\begin{figure}[ht]
\begin{center}
\includegraphics[width=0.47\textwidth]{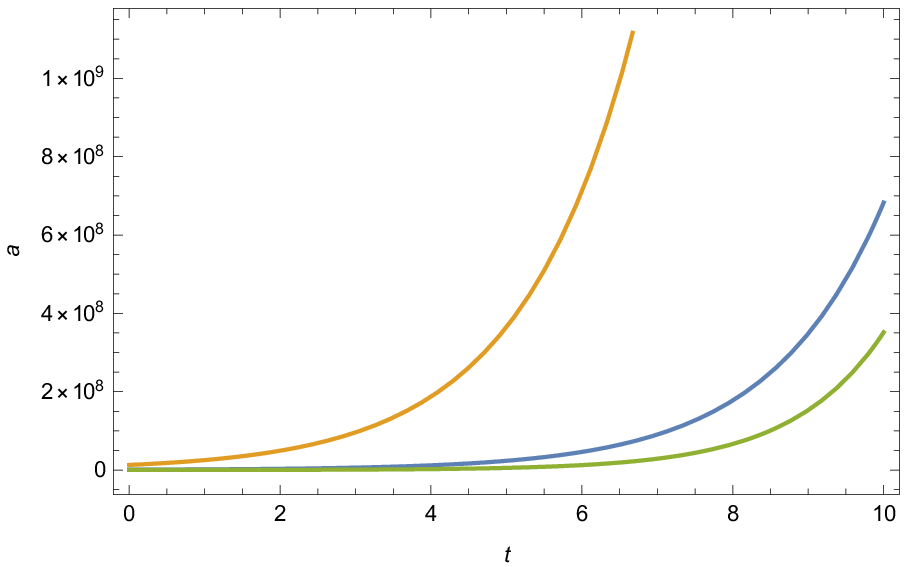}
\includegraphics[width=0.45\textwidth]{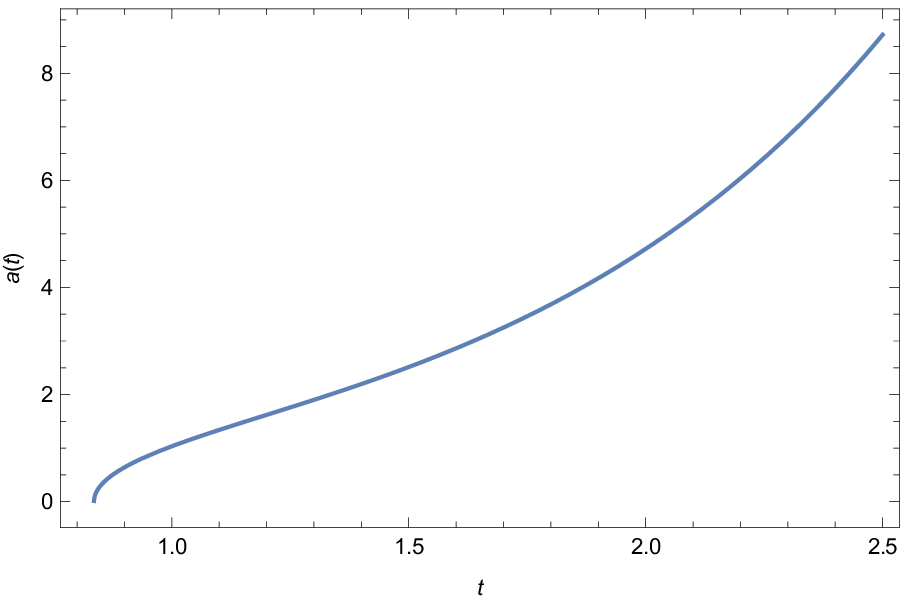}
\caption{Evolution of scale factor with respect to cosmic time. The various colors take on the following values of the index $n$ -- Yellow: $n = 3$, Blue: $n = 1.5$, and Green: $n = -0.5$.}
\label{mixed_pot_scal}
\end{center}
\end{figure}

\begin{figure}[ht]
\begin{center}
\includegraphics[width=0.35\textwidth]{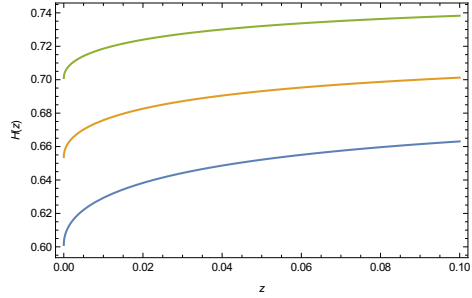}
\caption{Evolution of Hubble parameter with respect to redshift. The various colors take on the following values of the index $n$ -- Blue: $n = 2.5$, Yellow: $n = 3.5$, Green: $n = 4.5$. }
\label{power_law_hubb}
\end{center}
\end{figure}

\begin{figure}[ht]
\begin{center}
\includegraphics[width=0.35\textwidth]{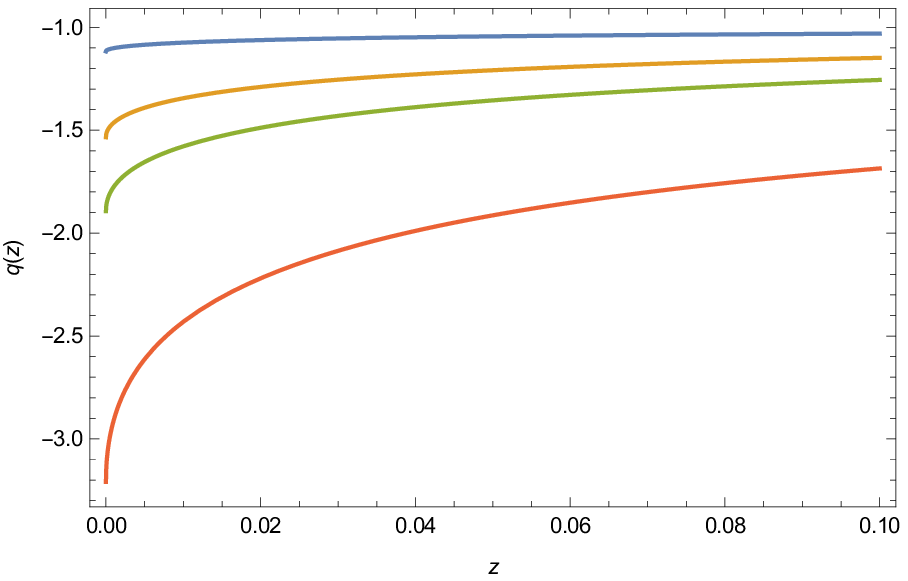}
\caption{Evolution of deceleration parameter with respect to redshift. The various colors take on the following values of the index $n$ -- Red $n = -0.5$, Green $n = 1.5$, Yellow $n = 3$, Blue $n = 5$.}
\label{power_law_dec}
\end{center}
\end{figure}

Comparing with the anharmonic oscillator in Eq.(\ref{gen}), we have $f_{1} = 3\frac{\dot{a}}{a}$, $f_{2} = 1$, $f_{3} = 1$. We study the integrability criterion and the point transformed version of the scalar field for this case. The integrability criterion yields an evolution equation for the scale factor which takes on the form
\begin{equation}
\frac{6}{(n+3)}\frac{\ddot{a}}{a} + \frac{12n}{(n+3)^2}\frac{\dot{a}^2}{a^2} - 1 = 0.
\end{equation}

\noindent A first integral can be found from this equation given as
\begin{equation}
\dot{a}^2 = \frac{(n+3)^2}{18(n+1)} a^2 + \lambda_{0} a^{-\frac{4n}{(n+3)}},
\end{equation}
which can straightforwardly be solved to give the scale factor
\begin{equation}
\label{exactscale2}
a(t) = \Bigg[\delta_{0} \cosh\Big(\sqrt{\frac{(1+n)}{2}}(t + 6 (3 + n) \delta_{1}\Big)\Bigg]^{\frac{(n+3)}{3(n+1)}},
\end{equation}
where $\delta_0,\, \delta_1$ are integration constants. In this instance all the values of $n$ (provided $n > - 3$) result in late-time accelerated expansion as can be seen in Fig.(\ref{mixed_pot_scal}).

Using the theorem, the evolution of the scalar field can be calculated by the solution of the point-transformed version of the anharmonic oscillator equation in Eq.(\ref{phigen}) giving
\begin{equation}
\phi(t) = -\frac{2n\sqrt{1-\cosh\Big(\sqrt{2(1+n)}(t + 6 (3 + n) \delta_{1}\Big)} \cosh\Big(\sqrt{2(1+n)}(t + 6 (3 + n) \delta_{1}\Big)}{3\sqrt{(1+n)}{_2}F{_1}\Big[\frac{1}{2},\frac{n}{3(1+n)};\frac{(3+4n)}{(3+3n)};\Big[\cosh\Big(\sqrt{\frac{(1+n)}{2}}(t + 6 (3 + n) \delta_{1}\Big)\Big]^2\Big]},
\end{equation}
where the solution appears in terms of hypergeometric functions. \medskip

To complete the picture, we also show the Hubble and deceleration parameters against redshift in Fig.(\ref{power_law_hubb}) and Fig.(\ref{power_law_dec}) respectively. The Hubble parameter does give the correct general behavior for most values of $n$ while the deceleration parameter depends more heavily on the precise value of the index.\medskip

The lagrangian functional form is found by taking the Friedmann equation in Eq.(\ref{fe1}) and solving it similar to the previous power-law instance. However, the differential equation that results turns out to be intractable and thus a solution is not found but at least the state parameters are found.

\subsubsection{Higgs Potential}
We now study a special case of a combination of power law potential, namely, the Higgs potential which is  given by
\begin{equation}
V(\phi) = V_{0} + \frac{1}{2} M^2 {\phi}^{2} + \frac{\lambda}{4}\phi^{4},
\end{equation}
such that $\frac{dV}{d\phi} = M^2 \phi + \lambda \phi^3$. Then the scalar field evolution equation looks like
\begin{equation}
\ddot{\phi} + 3\frac{\dot{a}}{a}\dot{\phi} + M^2 \phi + \lambda \phi^3 = 0.
\end{equation}

\begin{figure}[ht]
\begin{center}
\includegraphics[width=0.35\textwidth]{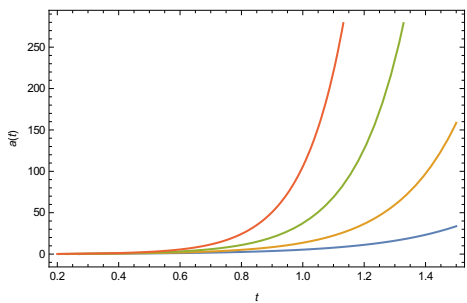}
\caption{Evolution of the scale factor with cosmic time. The various colors take on the following values of the mass term -- Blue: $M=1$, Yellow: $M=3$, Green: $M=5$, and Red: $M=7$.}
\label{higg_scale_fac}
\end{center}
\end{figure}

\begin{figure}[ht]
\begin{center}
\includegraphics[width=0.35\textwidth]{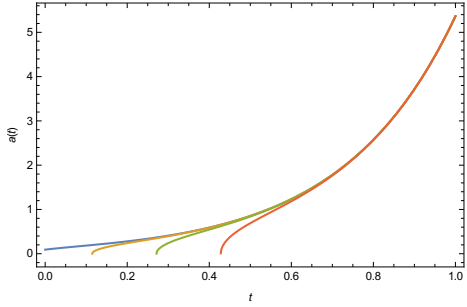}
\caption{Evolution of the scale factor cosmic time. The corresponding values of the $\lambda$ term are -- Blue: $\lambda=-20$, Yellow: $\lambda=-10$, Green: $\lambda=-0.1$, Red: $\lambda=0.1$, and Violet: $\lambda=1$. The mass term is kept fixed at $M=3$ for these cases.}
\label{higgs_scale_fac_lamb}
\end{center}
\end{figure}

Comparing with the anharmonic oscillator in Eq.(\ref{gen}), we have $f_{1} = 3\frac{\dot{a}}{a}$, $f_{2} = M^2$, $f_{3} = \lambda$. The integrability criterion and the point transformed version of the scalar field for this case yield the scale factor
\begin{equation}\label{exactscale4}
a(t) = \Bigg[\frac{1}{2 M^2} e^{\sqrt{2}Mt} - \lambda_{0} e^{-\sqrt{2}Mt}\Bigg]^{\frac{1}{2}}.
\end{equation}

The evolution of the scale factor with time is plotted for different choices of $M$ in Fig.(\ref{higg_scale_fac}). The plots all show accelerated expansion at late-times, however the rate of expansion varies depending on the choice of $M$. For increasingly large positive value of value for $M$, the rate of acceleration is increasingly fast for higher values. \medskip

The evolution of the scale factor with cosmic time is shown in Fig.(\ref{higgs_scale_fac_lamb}) for different values of the parameter $\lambda$. For all late-times of this scenario, the same highly accelerated setting follows for late-times. The behavior at early times does differ for the various values of $\lambda$ with slower expansion profiles. \medskip

The explicit expression for the scalar field can be determined by using the solution of the point-transformed version of the anharmonic oscillator equation following in Eq.(\ref{phigen}) which results in
\begin{equation}
\phi\left( t\right) =\phi_{0}\left[ C^{-1} \int \frac{1}{a(t) dt} -T_{0}\right] ^{-1} \frac{1}{a(t)}.
\end{equation}
Taking the scale factor in Eq.(\ref{exactscale4}) results in the scalar field
\begin{equation}
\phi(t) = D_{0} \frac{\sqrt{2} M}{\sqrt{\Big[4-\frac{2}{\lambda_{0} M^2}e^{2\sqrt{2}t}\Big]} {_2}F{_1}\Big[\frac{1}{4},\frac{1}{2};\frac{5}{4};\frac{e^{2\sqrt{2}t}}{2\lambda_{0} M^2}\Big]},
\end{equation}
where $D_{0}$ is a constant which consists of $\lambda$, $C$ and $\phi_{0}$. \medskip

\begin{figure}[ht]
\begin{center}
\includegraphics[width=0.35\textwidth]{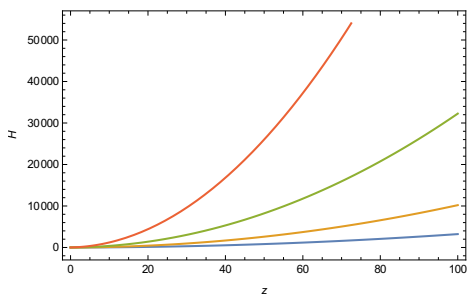}
\caption{Evolution of Hubble parameter with respect to redshift. The corresponding values of the mass term, $M$, are the same as in Fig.(\ref{higg_scale_fac}).}
\label{higgs_hubble}
\end{center}
\end{figure}

\begin{figure}[ht]
\begin{center}
\includegraphics[width=0.35\textwidth]{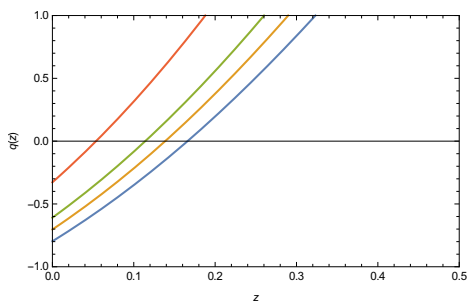}
\caption{Evolution of deceleration parameter with respect to redshift. The corresponding values of the mass term, $M$, are the same as in Fig.(\ref{higg_scale_fac}).}
\label{higgs_dec}
\end{center}
\end{figure}

We plot the cosmological parameters $H(z)$ and $q(z)$ as a function of redshift in Figs.(\ref{higgs_hubble},\ref{higgs_dec}). The Hubble parameter shows the correct general behavior, and differing behaviors for the various values of the mass term $M$. Similarly, the deceleration parameter tends to the correct present time value for certain mass term values $M$. This could constrain the acceptable values of $M$.

\subsection{Exponential Potential}
An exponential interaction potential is of particular importance in scalar field cosmology. Its found in higher-order and higher-dimensional gravitational theories \cite{Whitt:1984pd,0264-9381-10-2-009}, as well as nonperturbative settings with gaugino condensation \cite{deCarlos:1992kox}. Another important contribution to this kind of potential field is that shown in Ref.\cite{Halliwell:1986ja} where it was shown that a power-lie inflationary epoch is an attractor solution. \medskip

In sections $4.1$ and $4.2$, we discussed cosmological evolution where the self-interaction of the scalar field was given by a simple power law interaction or a suitable combination of power-law terms. The integrability condition was then exploited to determine a solution for the scalar field evolution. In the current setting, this approach cannot be used and so other methods will have to be employed to find a solution to the evolution equations. The exponential model under consideration here results in the evolution equation
\begin{equation}
\ddot{\phi} + 3\frac{\dot{a}}{a}\dot{\phi} + V_{0} e^{\alpha \phi} = 0,
\end{equation}
where $V=(V_0/\alpha)\, e^{\alpha \phi}$. \medskip

\noindent A very simple solution of this equation can be obtained giving
\begin{equation}\label{exactscale5}
a(t) = (t-t_{0})^{m^2},
\end{equation}
and the evolution of the scalar field can be written as $\phi(t) = -\frac{2}{\alpha} ln(t-t_{0})$. For a consistent solution one must enforce a restriction over the choice of $m^2$ as $m^2 = \frac{(2 + \alpha V_{0})}{6}$. \medskip

\begin{figure}[ht]
\begin{center}
\includegraphics[width=0.35\textwidth]{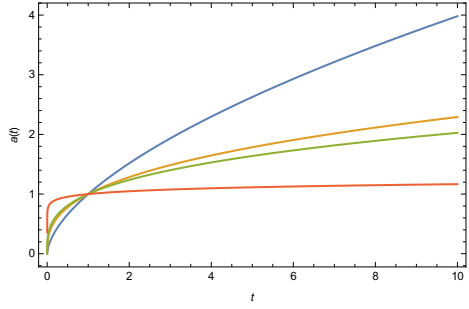}
\caption{Evolution of the scale factor with cosmic time. The various colors take on the following values of $V_0$ as follows -- Red: $V_{0} = 10$, Blue: $V_{0} = 1$, Yellow: $V_{0} = -1$, and Green: $V_{0} = -10$. The $\alpha$ parameter is fixed $\alpha = \frac{1}{6}$.}
\label{exp_scale_factor_V}
\end{center}
\end{figure}

\begin{figure}[ht]
\begin{center}
\includegraphics[width=0.45\textwidth]{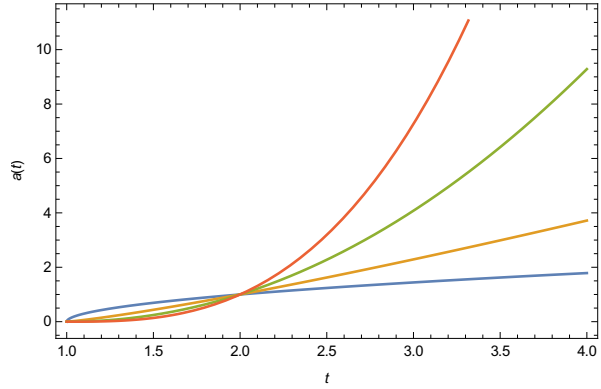}
\caption{Evolution of the scale factor with time for $V(\phi) = \frac{V_{0}}{\alpha} e^{\alpha \phi}$, with the value of a fixed $V_{0} = 1$, and varying $\alpha$ with the following values -- Red: $\alpha = 15$, Blue: $\alpha = 10$, Yellow: $\alpha = 5$, and Green: $\alpha = 1$.}
\label{exp_scale_factor_alpha}
\end{center}
\end{figure}

In Fig.(\ref{exp_scale_factor_V}) we plot the scale factor for several values of the constant $V_0$ with a representative value of the $\alpha$ parameter, $\alpha = \frac{1}{6}$. The most extreme, $V_{0} = 10$, shows relatively steady expansion, while the other curves gradually tend to level off with some turning inward and eventually turn out to be in decreasing expansion. In Fig.(\ref{exp_scale_factor_alpha}), we show the scale factor for several values of the parameter $\alpha$, with the steepest rate of expansion being found for $\alpha=15$. \medskip

Consider now the Friedmann equation in Eq.(\ref{fe1}), with the explicit form of the scalar field and potential in hand, the equations
\begin{equation}
f(H) + 6 H^2 - H\frac{df(H)}{dH} = \dot{\phi}^2 + 2V(\phi),
\end{equation}
becomes soluble. In fact, the solution in terms of the Hubble parameter turns out to be $f(H) = (6 - \lambda_{0})H^2 + \lambda_{1}H$, where $\lambda_{1}$ is a constant of integration and $\lambda_{0} = \Big(\frac{4}{m^{4}\alpha^2} + \frac{2V_{0}}{\alpha m^{4}}\Big)$. Therefore, the resulting functional form of $f(T)$ can be written as
\begin{equation}
f(T) = \Bigg(\frac{\lambda_{0}}{6}-1\Bigg)T + \lambda_{1} \Bigg(-\frac{T}{6}\Bigg)^{\frac{1}{2}}.
\end{equation}

\begin{figure}[ht]
\begin{center}
\includegraphics[width=0.35\textwidth]{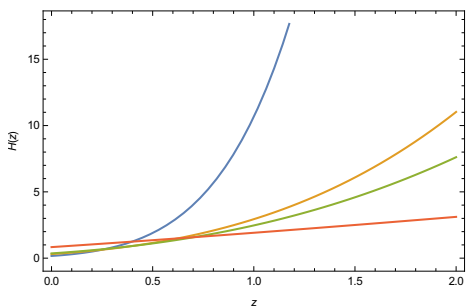}
\caption{Evolution of Hubble parameter with respect to redshift for different values of $\alpha$ -- Red: $\alpha = 3$, Green: $\alpha = 0.16$, Yellow: $\alpha = -0.16$, Blue: $\alpha = -1$.}
\label{exp_hubble}
\end{center}
\end{figure}

Again, as in the other potential functions being considered, we show the behavior of the Hubble and deceleration parameters against redshift. These are shown in Figs.(\ref{exp_hubble}).

\begin{figure}[ht]
\begin{center}
\includegraphics[width=0.35\textwidth]{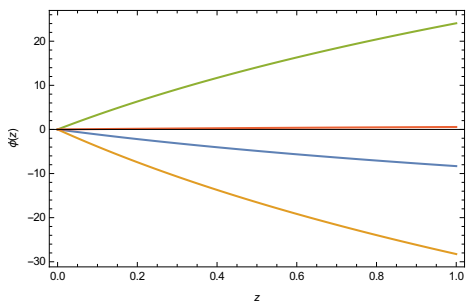}
\caption{Evolution of scalar field with respect to redshift for different values of $\alpha$ -- Red: $\alpha = 3$, Green: $\alpha = 0.16$, Yellow: $\alpha = -0.16$, Blue: $\alpha = -1$.}
\label{exp_decel}
\end{center}
\end{figure}

\subsection{Simple Logarithmic Potential}
Lastly, we study the effect of taking a logarithmic form of the scalar field interaction potential. These forms of potentials have shown promise in driving slow-roll inflation such as in Ref.\cite{Barrow:1995xb,Holden:1999hm}. For the current case, we consider the simple potential as
\begin{equation}
V(\phi) = \alpha ln \phi.
\end{equation}
This scenario gives rise to two sets of solutions when the Friedmann and scalar field evolution equations are considered.

\subsubsection{Case I}
Considering the Klein-Gordon equation for this potential we deduce the cosmic time evolution of the scalar field, which in turn leads to the explicit form of the scale factor. This is achieved using the Friedmann equation and gives 
\begin{equation}
a(t) = \alpha_{0} [e^{2C_1} - \alpha^2 t^2 - 2\alpha^2 C_2 t - \alpha^2 C_{2}^2]^{\frac{1}{6}},
\end{equation} 
and
\begin{equation}
\phi(t) = \frac{1}{\alpha^{\frac{1}{2}}} [e^{2C_1} - \alpha^2 t^2 - 2\alpha^2 C_2 t - \alpha^2 C_{2}^2]^{\frac{1}{2}}.
\end{equation}

\begin{figure}[ht]
\begin{center}
\includegraphics[width=0.35\textwidth]{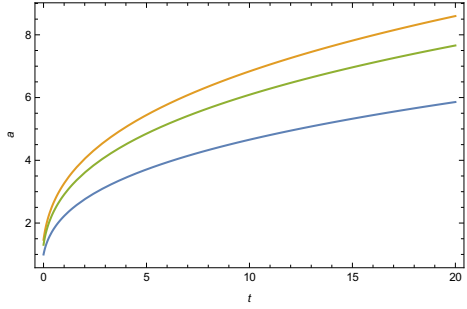}
\caption{Evolution of scale factor with respect to cosmic time. The various colors take on the following values of $\alpha$ -- Blue: $\alpha = -100$, Green: $\alpha = -500$, Yellow: $\alpha = -1000$.}
\label{log_scale_fac}
\end{center}
\end{figure}

We plot the evolution of the scale factor with respect to cosmic time for different initial conditions in Fig.(\ref{log_scale_fac}). It is clear from the figure that the scale factor initially expands rapidly, however, this eventually tends off and slows down. Therefore, this particular case does not produce the correct expansion history we observe physically. \medskip

While an analytic solution can be found for $f(T)$ in this case, it turns out to be very long and given that the solution does not reproduce the correct expansion history we omit it here.
 
\subsubsection{Case II}
In the second solution to the Klein-Gordon equation we find that the scale factor turns out to take the form
\begin{equation}
a(t) = e^{\frac{-(t-d)^{-2m} [(2mdt - mt^2 +d^2 (-m+(1-\frac{t}{d})^{2m})) V_{0} + 2(m-1)^2 m^2 (t-d)^{2m} ln(1-\frac{t}{d})]}{6m^2 (1-m)}},
\end{equation} 
while the scalar field is given by
\begin{equation}
\phi(t) = (t-d)^m .
\end{equation}

\begin{figure}[ht]
\begin{center}
\includegraphics[width=0.35\textwidth]{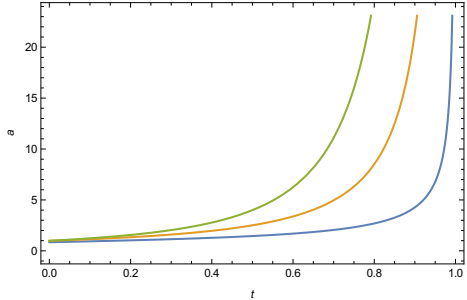}
\caption{Evolution of scale factor with respect to cosmic time. The various colors take on the following values of $m$ -- Blue: $m = 3$, Green: $m = 0.5$, Yellow: $m = -1.5$.}
\label{log_scale_fac2}
\end{center}
\end{figure}

Unlike in the first instance, the scale factor clearly shows late-time accelerated expansion in the current setting, for various initial conditions, as shown in Fig.(\ref{log_scale_fac2}). Again, as in the first case, the resulting $f(T)$ solution turns out to be very long, with many contributory terms, and so it is omitted.

\section{Conclusion}
In this work we consider generalized teleparallel gravity through the prism of a scalar field source that is minimally coupled to the gravitational lagrangian. We calculate the evolution equation for this relation and use this to determine different state parameters and the generalized lagrangian function for various potentials. This is achieved using the very useful theorem presented in section 3. As shown this can have very useful cosmological applications for different epochs of the history of the universe. \medskip

In particular, we first consider a power law potential for the scalar field which straightforwardly leads to a power law scale factor using the conditional constraint coming from the theorem. With this in hand we were able to simply determine the Hubble and deceleration parameters. However, our interest stretched to the EoS parameter for the scalar field which, in some instances, achieved the $\omega=-1$ observational constraint. This parameter depends on both the solution for the scalar field itself as well as the assumption on its potential, as shown through Eq.(\ref{EoS_eqn}). Finally, we used the scalar field and scale factor solutions to determine the generalized lagrangian function, $f(T)$. The solution that was found is interesting because power law solutions have so far been limited to one term, as in Ref.\cite{Bengochea:2008gz}, while in our case we find that generic two term power law instances may also have important effects in the cosmological regime of teleparallel gravity. \medskip

In our second considered potential we take a combination of power laws with one of them being a quadratic. This is chosen so that the theorem can be utilized in this instance. Through the theorem constraint equation we straightforwardly determine the scale factor. In this case, it turns out to be more complex, exhibiting behavior which mimics a power law at late-times. At least early times some of the possible values of the power index, $n$, gives a steep increase in the growth of the universe possibly imitating some aspects of the inflationary epoch. Using the result for the scale factor we then determine the scalar field itself which turns out to also be a complex composition of hypergeometric functions. This last point holds further progress toward the general lagrangian function which remains undetermined analytically. As a sub-class we consider the Higgs potential instance which has a specific form for the general power law for the potential in question. However, in this case we also find a scalar field dependent on hypergeometric functions which continues to hold back a full determination of the gravitational lagrangian. On the other hand, we do find the state parameter set for both cases. \medskip

To better investigate the full breath of possible forms of the scalar field potential we finally investigate the cases of exponential and logarithmic instances. In the exponential case, we also find a power law scale factor emerging. This aides in the later determination of the lagrangian function which turns out to be a combination of the TEGR term with a new square root term appearing. For the logarithmic case two solutions emerge from the theorem constraint equation which give results for the scale factor that are quite different. These give roughly, either power-law or logarithmic style scale factor evolution. In the second case, the scalar field gives a power law evolution in time. However, as with the combined potential instance, the Friedmann equation for the function, $f(T)$, turns out to be intractable and so the full analysis is not possible analytically. \medskip

We have presented some possible cosmic evolution parameters for various potentials for a minimally coupled scalar field. These have confirmed some familiar solutions for the unknown, $f(T)$, lagrangian while also presenting some new terms. It would be interesting to determine what other cosmological effects these lagrangians would have but this is outside the scope of this work.    \medskip

\section{Acknowledgment}
The authors would like to thank Professor Narayan Banerjee for useful comments and suggestions. SC was supported by the National Post-Doctoral Fellowship ($File Number: PDF/2017/000750$) from the Science and
Engineering Research Board (SERB), Government of India.

\bibliographystyle{unsrt}
\bibliography{references}

\end{document}